\documentclass[ams,showpacs,nofootinbib,preprint]{revtex4}
\usepackage{amsmath}
\usepackage{epsfig}
\usepackage{endnotes}
\let\footnote=\endnote

\newcommand{\be}{\begin{equation}}
\newcommand{\ee}{\end{equation}}

\newcommand{\ber}{\begin{eqnarray}}
\newcommand{\eer}{\end{eqnarray}}

\begin{document}

\title{Frame dependence of $^3$He transverse $(e,e')$ response functions at intermediate momentum transfers }

\author{Victor D. Efros$^{1}$,
  Winfried Leidemann$^{2,3}$, 
  Giuseppina Orlandini$^{2,3}$,
  and Edward L. Tomusiak$^{4}$ 
  }

\affiliation{
  $^{1}$Russian Research Centre 
  "Kurchatov Institute",  123182 Moscow,  Russia\\ 
  $^{2}$Dipartimento di Fisica, Universit\`a di Trento, I-38123 Trento, Italy \\
  $^{3}$ Istituto Nazionale di Fisica Nucleare, Gruppo Collegato di Trento,
  I-38123 Trento, Italy \\
  $^{4}$Department of Physics and Astronomy,
  University of Victoria, Victoria, BC V8P 1A1, Canada\\
}

\begin{abstract}
The transverse electron scattering response function of $^3$He was recently
studied by us in the quasi-elastic peak region for momentum transfers $q$ between
500 and 700 MeV/c. Those results, obtained using the Active Nucleon
Breit frame (ANB), are here supplemented by calculations in
the laboratory, Breit and ANB frames using the two-fragment model discussed
in our earlier work on the frame dependence of the the longitudinal
response function $R_L(q,\omega)$.  We find relatively frame
independent results and good agreement with experiment especially 
for the lower momentum transfers. This agreement occurs when we
neglect an $\omega$-dependent piece of the one-body current
relativistic correction.  An inclusion of this term leads however to a rather
pronounced frame dependence at $q=700$ MeV/c. A discussion of this term is given here.
This report also includes a correction to our previous
ANB results for $R_T(q,\omega)$.
\end{abstract}

\bigskip


\maketitle

In a recent publication Ref.~\cite{ELOT10}
we presented calculations of $R_T(q,\omega)$, the
transverse electron scattering response function of $^3$He. That calculation
was based on the Lorentz integral transform (LIT) technique \cite{ELO94,ELOT04,ELOB07}
and included the AV18 \cite{AV18} NN + UrbanaIX \cite{UrbIX} NNN 
potentials, the lowest order
relativistic corrections to the one-body current and consistent
two-body currents (MEC) due to meson
exchange. Detailed references to these aspects can be found in \cite{ELOT10}.
The laboratory (lab) frame response function was calculated by first computing in the
Active Nucleon Breit (ANB) frame and then transforming to the lab frame.
Our earlier work \cite{ELOT05} on the frame dependence of the longitudinal
response function $R_L(q,\omega)$ showed that the ANB frame appeared to
diminish the effects of using non-relativistic nuclear dynamics. In
addition, a two-fragment model was introduced which
was found to significantly reduce frame dependence. In that model one assumes
a quasi-elastic knock-out of a nucleon such that the residual nucleus remains in
its lowest energy state. The relative momentum of the two fragments
is determined in a relativistically correct way and the energy that is used as 
input to the non-relativistic  calculation is obtained from that momentum by the 
usual non-relativistic relation. Considering for example the lab system, the quasi-elastically
knocked out nucleon has momentum $q$, thus a difference between non-relativistic and 
relativistic kinetic energies of $\Delta T_{\rm 1b} = q^2/2M - (M^2 + q^2)^{1/2} +M$  ($M$ is nucleon mass). 
Taking for example $q=700$ MeV/c one has $\Delta T_{\rm 1b}=29$ MeV. 
This relativistic effect is automatically taken into account in the two-fragment model. 
It is important to state that this model only fixes a kinematical input 
while the full three-nucleon dynamics is still treated completely via the
LIT technique. Thus a comparison of calculations done in the ANB frame with those
done in any other frame (including the ANB frame) but with the
two-fragment model provides some indication
of uncertainties due to non-covariance.

In dealing with the transverse response function of $^3$He near the quasi-elastic peak
where the energy-momentum of the virtual photon is absorbed predominantly by the single
ejected nucleon one can again adopt the two-fragment model. Compared to the longitudinal
response function there is however a slight difference in applying the two-fragment model
to the transverse response. For the former we did not include a two-body charge operator,
while for the latter we do consider, as mentioned above, an MEC. 
It is not evident that such a two-body current should
be taken into account in the two-fragment model, since it violates the assumption
of a quasi-elastic knock-out. Instead one may assume the following scenario (lab-system): 
initially two nucleons have opposite and equal momenta ${\bf p}$ and $-{\bf p}$ while
the third nucleon is at rest. If the photon momentum is transferred to the 1-2 pair 
then their momenta in the final state become ${\bf p}+{\bf q}/2$ and $-{\bf p}+{\bf q}/2$. 
Setting the total kinetic energy energy of the 1-2 pair equal to $q^2/2M$, i.e. the excitation  energy of the quasi-elastic peak, one can solve for ${\bf p}$  and hence the kinetic energy of the 1-2 pair using either relativistic or non-relativistic kinematics. If one takes $q=700$ MeV/c then one can show this difference varies between 16 MeV (${\bf p}$ and ${\bf q}$ orthogonal) and
29 MeV (${\bf }p$ and ${\bf q}$ parallel).
As long as MEC matrix elements are small their main contribution arises from interference with the one-body current operator. Thus most probably their most significant effect is in the forward direction (${\bf p}$ and ${\bf q}$ parallel) which is the assumption of the two-fragment model.
 If, however,
MEC and one-body current contributions are similar in size, a part of the MEC contribution 
is not taken into account with the proper final-state energy in the two-fragment model. Even then
one may assume in most cases that the MEC changes only mildly with energy and thus errors
remain small. 
Moreover, as already pointed out in \cite{ELOT10} MEC have only a minor effect in the peak region.

In principle one should also consider implicit MEC contributions, which arise for example by
application of the Siegert theorem.
Though here we do not make use of the classical
Siegert operators we need to better investigate the actual form of our one-body current. It 
is described in Eqs.~(11) to (16) of Ref.~\cite{ELOT10}. One of the pieces of this
current, referred to here as the $\omega$-dependent piece, appears
as $(\omega/M){\bf j}_\omega$ where $\omega$ is the energy transfered to
the nuclear system by the virtual photon and where 
\begin{equation}
{\bf j}_\omega=\ e^{i{\bf qr}'}\frac{G_E-2G_M}{8M}
\left({\bf q}+i\kappa [{\vec \sigma}\times{\bf q}]+2i[{\vec \sigma}\times{\bf p}']
\right).
\end{equation}
Here $\kappa$ is given by
\begin{equation}
\kappa=1+2P_i/Aq\ ,
\end{equation}
${\bf r}'$ and ${\bf p}'$ are relative coordinates and momenta of a single particle,
P$_i$ is the magnitude of the target momentum, and $A$ is the
nucleon number of the target. This current contains an implicit MEC contribution.
For example for deuteron photodisintegration in \cite{WLA} it was shown
that the contribution due to the spin-orbit term is almost exclusively an
MEC contribution. For quasi-elastic kinematics, however, one has a completely different
situation. 
In this respect it is helpful to realize that the operator form of the spin-orbit current, i.e.
${\vec \sigma}\times{\bf q}$, is identical to that of the non-relativistic spin current and thus, in the quasi-elastic region, should lead like the latter mainly to one-body knock-out in forward direction. 

For the explicit MEC we have studied the interference with the one-body current.
As to be expected we find a rather weak interference effect only. Nonetheless it is the interference, 
not the MEC by itself, which leads to the bulk of the MEC contribution. 
Thus we may conclude that for quasi-elastic kinematics MEC contributions are quite small and due to 
interference with the one-body current and hence can be safely
included in the two-fragment model.

In Fig.~1 we show various results for $R_T$ in the ANB frame.
These results are obtained by calculating in the ANB frame and then transforming to
the lab frame.  The effect of ${\bf j}_\omega$ is negligibly small in the ANB frame
since $\kappa$=0 in this case and $\omega$=0 at the quasi-elastic peak, thus its
contribution is not shown separately. The ANB frame results in the left-hand panels of Fig.~1 
replace those of
\cite{ELOT10} where an error resulting from an incorrect form
factor argument gave results higher than experiment.  In the
present corrected version the ANB results are now lower than
experiment. In order to show that the difference does not result
from an insufficient consideration of higher multipole contributions
we show in Fig.~2 convergence tests for electric and magnetic LIT strength
for both isospin channels as a function of final state multipolarity at q=700 MeV/c.  It is clear that the results 
are essentially fully converged
by J$_f$=35/2. As an additional accuracy test we calculated the non-energy weigthed sum rules for $R_T(q,\omega)$
with the non-relativistic one-body current operator. Comparing with the well known
non-energy weighted spin sum rule \cite{OrT} we obtain 98.4\%, 98.5\%, and 98.0\% at $q=500$, 600, and 700 MeV/c, 
respectively. It shows that the sum rule is quite well satisfied.

It is of interest to compare this ANB calculation with
one where the ANB calculation uses the two-fragment approximation for setting
the kinematics. The results of this 
are depicted in the right-hand panels of Fig.~1  where one notes some difference
between the two concerning the peak height only. 
Indeed whereas in other frames the two-fragment model
is effective in obtaining the quasi-elastic peak in the correct position 
this occurs automatically in the ANB frame without the need for the
two-fragment model. On the other hand the proper treatment of relativistic kinematics
leads to a slight increase of the peak heights, namely by 3.3\%, 4.6\%, and 6.1\% at
$q=500$, 600, and 700 MeV/c, respectively. Such an increase improves the agreement
with experimental data, however, the theoretical peak heights are still somewhat too low,
but one should also take into account that some additional strength could come from currents involving the $\Delta$-resonance.

Our previous work \cite{ELOT05} on frame dependence in $R_L(q,\omega)$ compared
ANB frame results with those obtained in the lab and Breit frames all calculated using
the two-fragment model. 
In Fig.~3 we show the corresponding results for $R_T(q,\omega)$ (MEC and ${\bf j_\omega}$ 
left out). As expected the experimental quasi-elastic peak position is reproduced
in all three cases. For the quasi-elastic peak height one finds differences
between 7\% ($q$=500 MeV/c) and 10\% ($q$=700 MeV/c) showing an existing but not excessive
frame dependence, which is about twice as large as the experimental error. The range
of theoretical results covers the experimental data in the peak region.

Our conclusion concerning the relatively mild frame dependence relies on the assumption that  
contributions due to MEC and ${\bf j_\omega}$ remain small in other frames. While this 
assumption seems to be rather safe for the MEC it could be not correct for ${\bf j_\omega}$, 
since it depends on the frame dependent variables $\kappa$ and total energy transfer $\omega$.
To better investigate this question we show in
Fig.~4 the contributions of the various currents
for the lab frame calculation using the two-fragment model. One can see that ${\bf j}_\omega$ has 
a significant effect on the lab system calculations
and that the total reduction due to relativistic effects becomes very large (37\% at $q$=700 MeV/c). 
Comparing the result with the ${\bf j_\omega}$ contribution to the corresponding ANB frame results 
in the right-hand panels of Fig.~1,
one finds reductions of the peak heights of about
5\% ($q$=500 MeV/c), 9\% ($q$=600 MeV/c), and 16\% ($q$=700 MeV/c). Thus one has a rather pronounced 
frame dependence at $q$=700 MeV/c. At this point we should remind the reader that the two-fragment model 
takes into account relativistic effects for the kinematics only, and
that dynamical relativistic effects are not considered at all. 

In such a situation with rather frame dependent results one should choose the frame which leads to the 
smallest relativistic corrections. In \cite{ELOT05} we already pointed that the ANB frame is preferable 
in case of quasi-elastic kinematics, since in this frame the nucleon momenta in initial and final states 
are only of the order of $q/2$, while in other frames one finds larger momenta, e.g., in the lab frame the 
knocked out nucleon has momentum $q$. In fact the reduction of $R_T$ due to relativistic effects is 
relatively small in the ANB frame (13\% instead of the 37\% in the lab frame at $q$=700 MeV/c). 
Therefore we consider the ANB results
to be more realistic than, e.g.,  the lab results. On the other hand it would be desirable to further 
reduce the frame dependence. This is possibly achieved by taking into account boost corrections.
This could be done in a way similar
to that used in deuteron electrodisintegration \cite{BA,RGWA}.

Acknowledgments of financial support are given to
the RFBR, grant 10-02-00718 and RMES, grant NS-7235.2010.2 (V.D.E.),
and to the National Science and Engineering Research Council of Canada (E.L.T.).

\vskip 1.0 in

\begin{figure}[ht]
\centerline{\resizebox*{14.cm}{19.cm}{\includegraphics[angle=0]{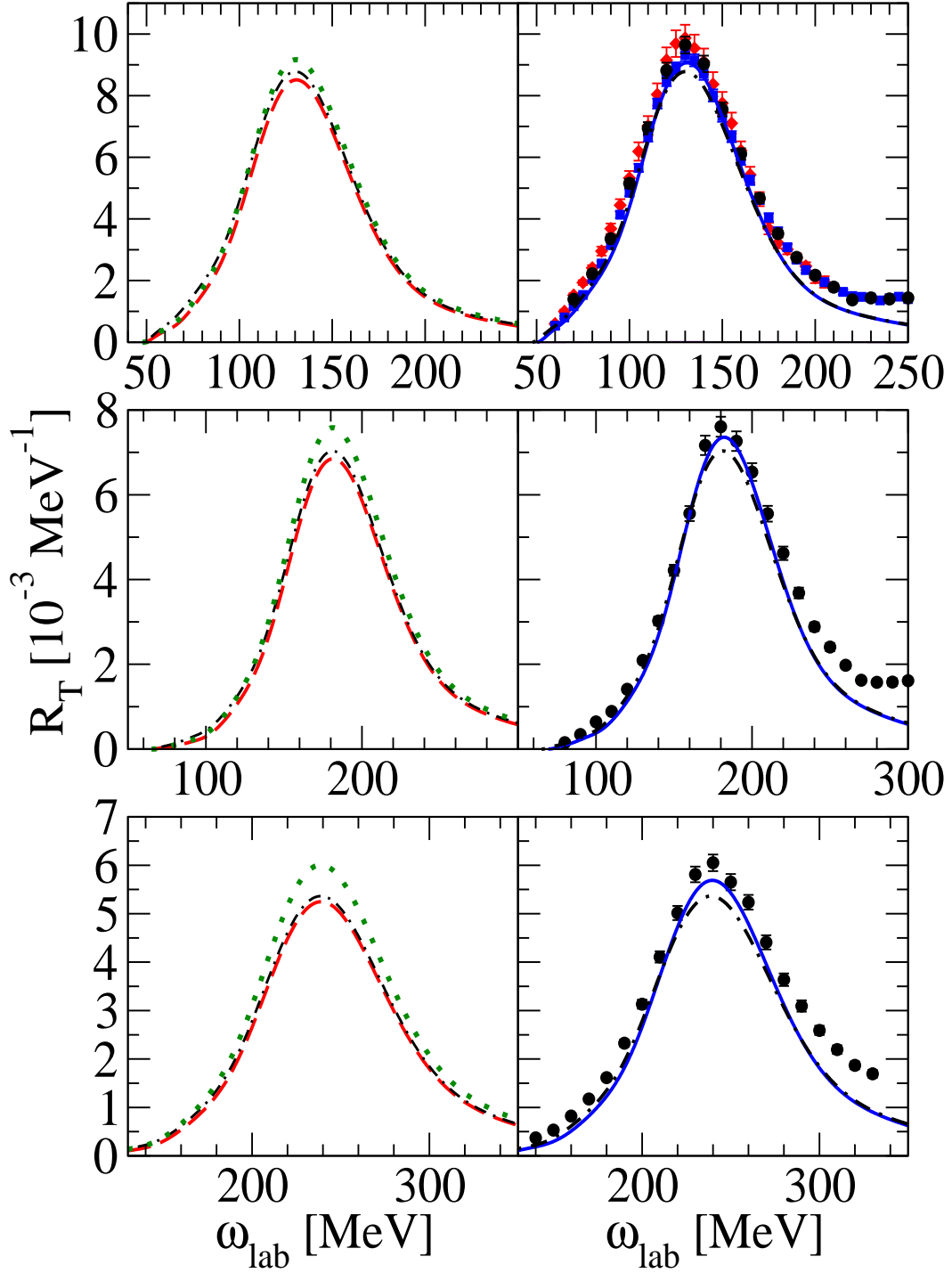}}}
\caption{$R_T(q_{\rm lab},\omega_{\rm lab})$ at $q_{\rm lab}=500,$ 600, and 700 MeV/c from ANB frame calculation. In left-hand panels results without use of two-fragment model with non-relativistic one-body current (dotted), relativistic one-body current (dashed), and
relativistic one-body current + MEC (dash-dotted). In right-hand panels results with relativistic one-body current + MEC without (dash-dotted) and with (solid) use of two-fragment model.
Experimental data from \cite{Saclay} (squares),  \cite{Bates} (diamonds),
\cite{world} (circles).  }
\label{Fig.1}
\end{figure}

\begin{figure}[ht]
\centerline{\resizebox*{15.cm}{18.cm}{\includegraphics*[angle=0]{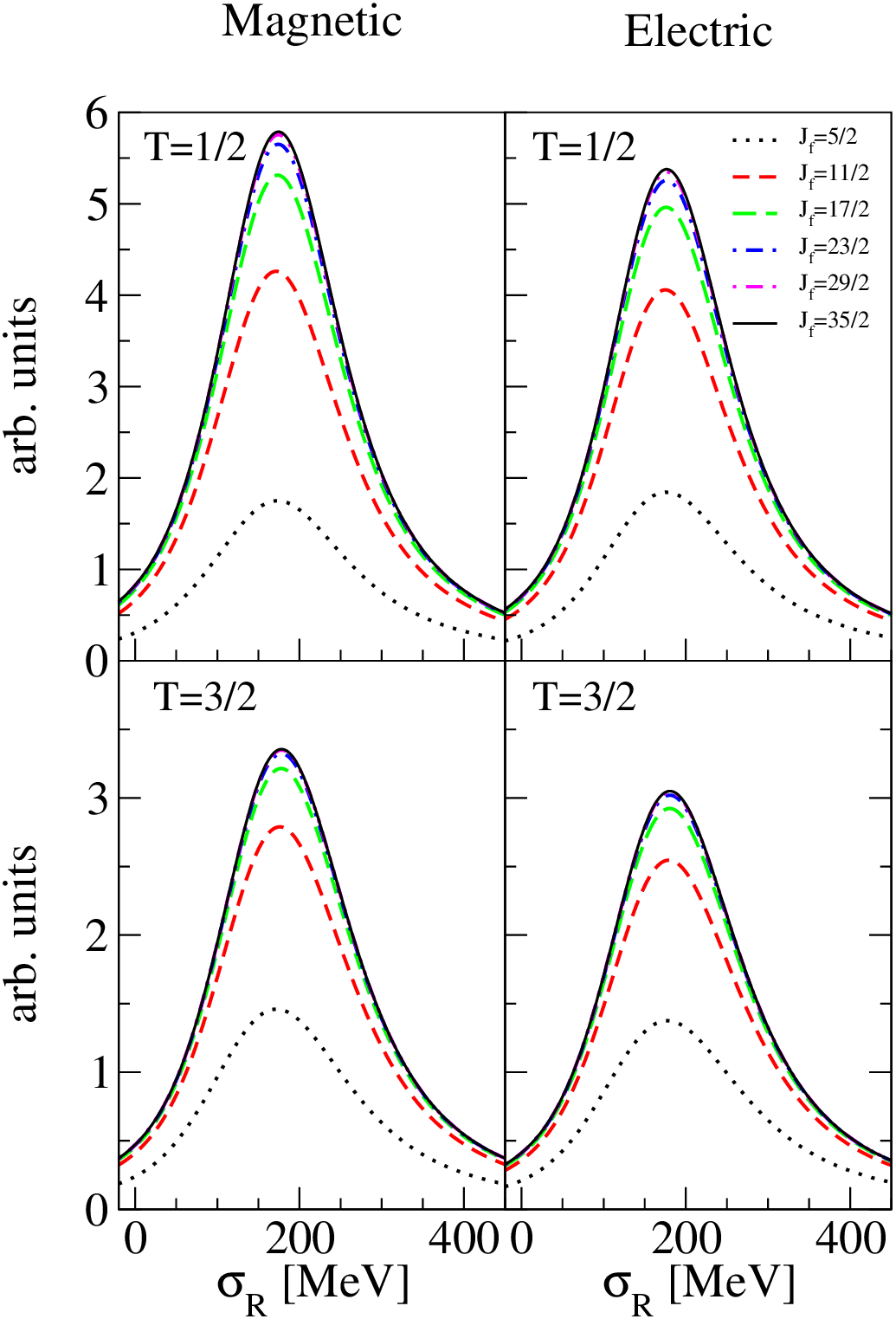}}}
\caption{Convergence of Lorentz integral transforms with maximal final state angular quantum number
$J_f$ for magnetic (left panels)
and electric (right panels) transition strength with final state isospin quantum number 
$T_f$=1/2 (upper panels) and $T_f$=3/2 (lower panels); the LIT resolution parameter $\sigma_I$ \cite{ELOB07} is equal to 50 MeV.}
\end{figure}

\begin{figure}[ht]
\centerline{\resizebox*{15.cm}{20.cm}{\includegraphics*[angle=0]{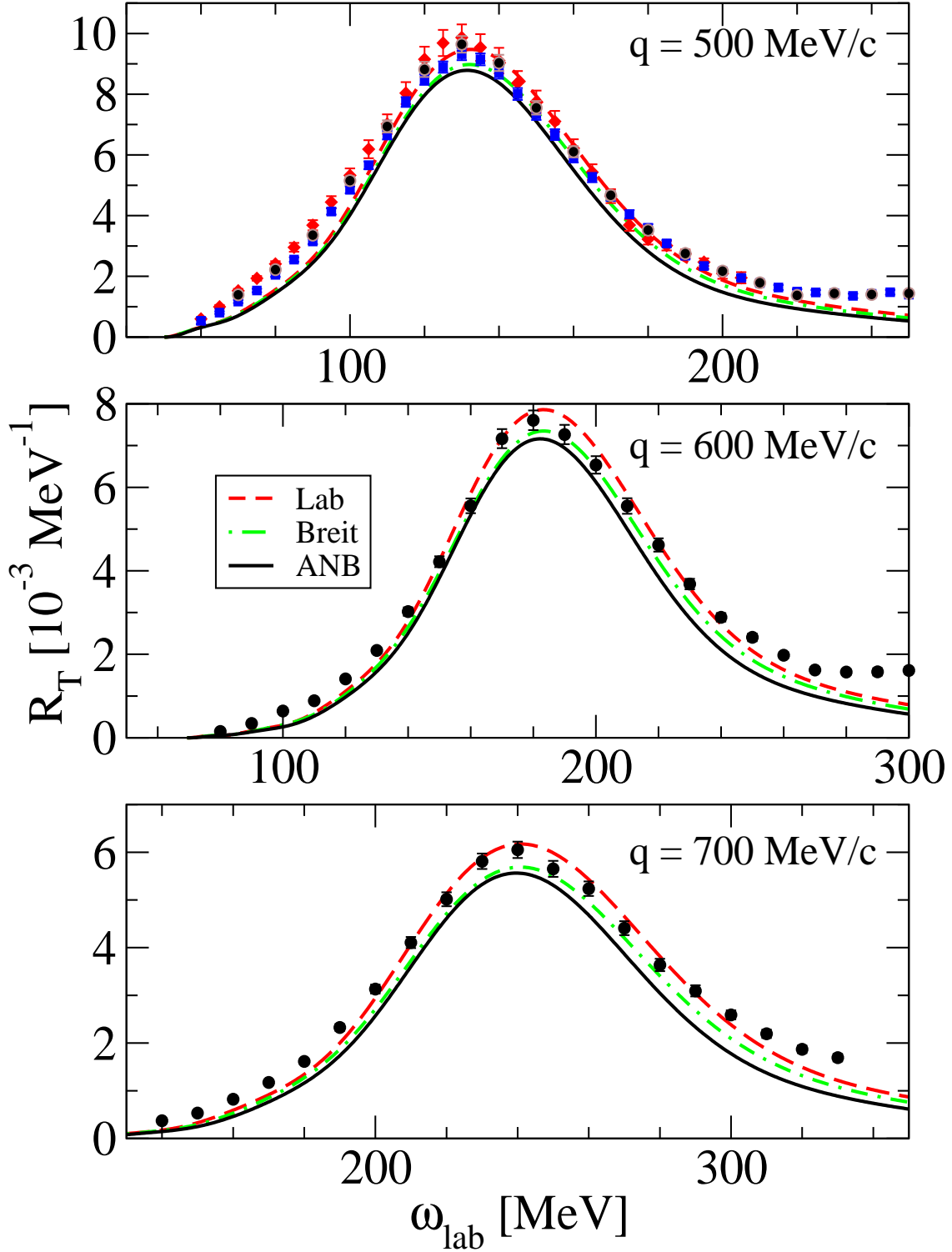}}}
\caption{$R_T(q_{\rm lab},\omega_{\rm lab})$ at $q_{\rm lab}=500,$ 600, and 700 MeV/c  with relativistic one-body current less $j_\omega$ with use of two-fragment model from calculations in various frames, namely ANB (solid), lab (dashed), and Breit (dot-dashed) frames.
Data as in Fig.~1.}
\end{figure}

\begin{figure}[ht]
\centerline{\resizebox*{15.cm}{20.cm}{\includegraphics*[angle=0]{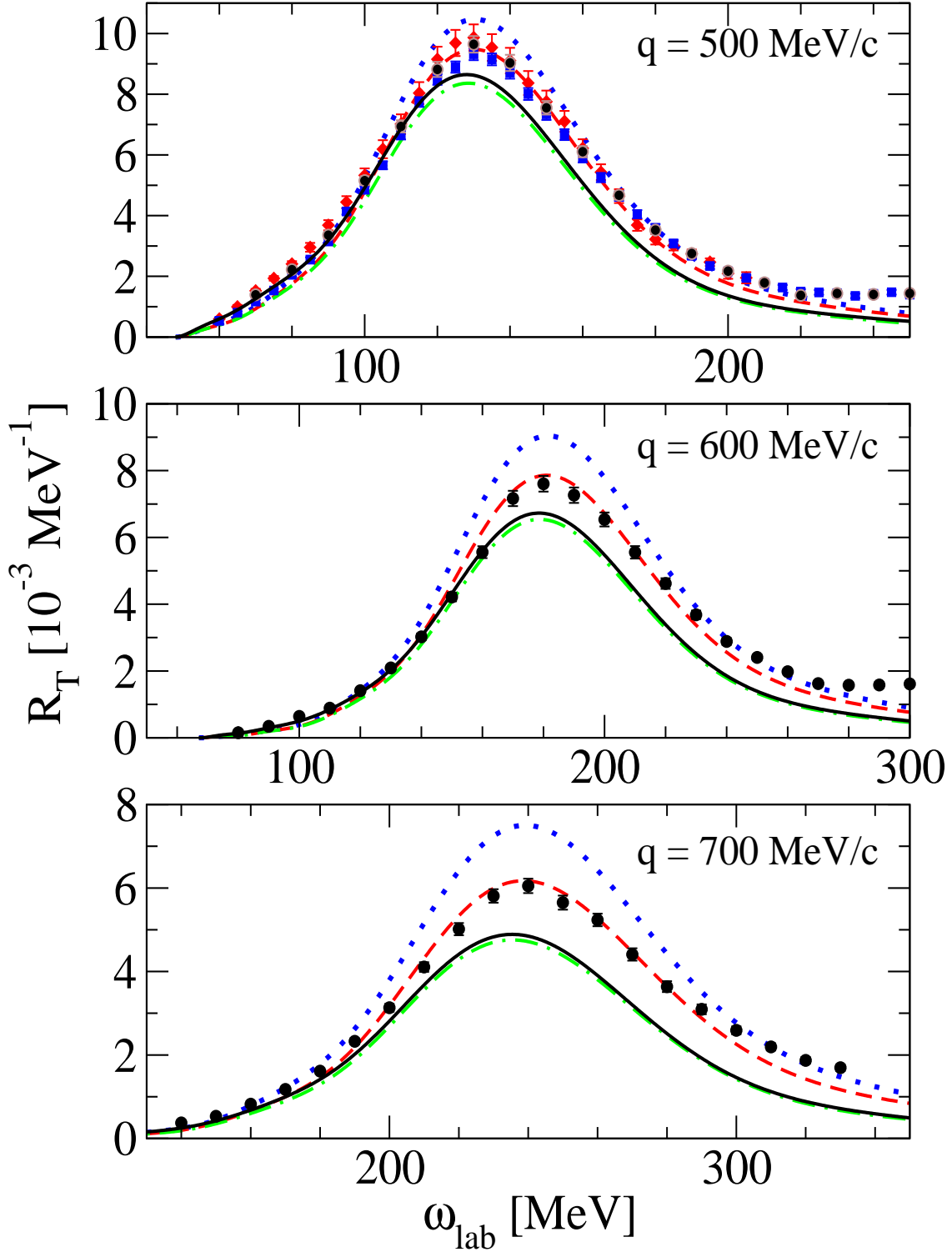}}}
\caption{$R_T(q_{\rm lab},\omega_{\rm lab})$ at $q_{\rm lab}=500,$ 600, and 700 MeV/c from lab frame calculation with use of two-fragment model: non-relativistic one-body current (dotted), relativistic one-body current less $j_\omega$
(dashed), relativistic one-body current (dash-dotted), and relativistic one-body current + MEC
(solid). Data as in Fig.~1.}
\end{figure}

\end{document}